\documentclass{cpbtex}

\usepackage{graphicx}
\usepackage{float}
\usepackage{placeins}
\usepackage[font=footnotesize,labelfont=bf,labelsep=period]{caption}

\graphicspath{{figures/}}
\newcommand{\TwoColumnWideAbstract}{}

\newcommand{\Tr}{\operatorname{Tr}}
\newcommand{\dd}{\mathrm{d}}
\newcommand{\ii}{\mathrm{i}}
\newcommand{\e}{\mathrm{e}}
\newcommand{\id}{\mathbb{I}}
\newcommand{\cH}{\mathcal{H}}
\newcommand{\cO}{\mathcal{O}}
\newcommand{\norm}[1]{\left\lVert #1\right\rVert}
\newcommand{\abs}[1]{\left\lvert #1\right\rvert}
\newcommand{\QCQ}{\mathrm{QCQ}}
\newcommand{\paperabstract}{%
High-fidelity CZ gates are central to superconducting quantum processors, but implementations based on flux tuning are still sensitive to pulse distortion. Conventional predistortion, typically used for an isolated gate, can recover the desired flux at the chip but it fails if the flux line memory exists, causing the fidelity of repeated CZ gates to drop rapidly. To address this issue, we model the flux line distortion as the dynamics of a stateful classical actuator coupled to a quantum system. Using a first-order Dyson expansion, we derive the error generators induced by variations in the initial flux-line state. We then design a robust CZ gate by optimizing the flux pulse to suppress these generators and minimize the residual flux line state at the gate exit. A one-pole flux line model shows the expected first-order robustness plateau. For a more practical three-pole model, the optimized CZ pulse achieves $F_{\mathrm{avg}}=99.998\%$, suppresses all first-order error generators, and brings the residual flux line state close to zero. With no additional waiting time between gates, our robust pulse achieves $F_{\mathrm{avg}}=99.97\%$ for the complete ten-gate sequence and reduces the sequence infidelity by a factor of about $2.3\times10^{3}$ relative to the baseline under the same predistortion protocol. These results show that explicitly accounting for flux line memory maintains high-fidelity CZ operation across repeated gate sequences and addresses a key limitation of conventional predistortion.
}

\newcommand{\paperkeywords}{superconducting quantum processors,
robust CZ gate, flux line distortion, robust quantum control}

\makeatletter
\renewcommand{\thesection}{\arabic{section}.}
\renewcommand{\thesubsection}{\arabic{section}.\arabic{subsection}.}
\renewcommand{\thesubsubsection}{\arabic{section}.\arabic{subsection}.\arabic{subsubsection}.}
\renewcommand\section{\@startsection{section}{1}{\z@}%
  {-3.0ex \@plus -1ex \@minus -.2ex}%
  {1.5ex \@plus .2ex}%
  {\normalfont\large\bfseries\raggedright}}
\renewcommand\subsection{\@startsection{subsection}{2}{\z@}%
  {-2.5ex \@plus -1ex \@minus -.2ex}%
  {1.0ex \@plus .2ex}%
  {\normalfont\normalsize\bfseries\raggedright}}
\renewcommand\subsubsection{\@startsection{subsubsection}{3}{\z@}%
  {-2.0ex \@plus -1ex \@minus -.2ex}%
  {.8ex \@plus .2ex}%
  {\normalfont\normalsize\bfseries\raggedright}}
\newcommand{\cleansectionlabel}[1]{%
  \begingroup
    \edef\@currentlabel{\arabic{section}}%
    \label{#1}%
  \endgroup
}
\newcommand{\cleansubsectionlabel}[1]{%
  \begingroup
    \edef\@currentlabel{\arabic{section}.\arabic{subsection}}%
    \label{#1}%
  \endgroup
}
\newcommand{\cleanappendixlabel}[1]{%
  \begingroup
    \edef\@currentlabel{\Alph{section}}%
    \label{#1}%
  \endgroup
}
\newcommand{\makewidefrontmatter}{%
  \twocolumn[{%
    \begin{@twocolumnfalse}
      \maketitle
      \begin{abstract}
        \paperabstract
      \end{abstract}
      \noindent\textbf{Keywords:} \paperkeywords

      \vspace{1.2em}
    \end{@twocolumnfalse}
  }]%
}
\makeatother

\begin{document}

\title{Robust CZ gate against flux line memory}
\author{Yao Song$^{1}$ and Xiu-Hao Deng$^{2,3}$%
\thanks{Corresponding author. E-mail:~dengxiuhao@iqasz.cn}\\
$^{1}$Shenzhen Institute for Quantum Science and Engineering,\\
Southern University of Science and Technology, Shenzhen 518048, China\\
$^{2}$International Quantum Academy, Shenzhen 518048, China\\
$^{3}$Shenzhen Branch, Hefei National Laboratory, Shenzhen 518048, China}
\date{}
\ifdefined\TwoColumnWideAbstract
  \makewidefrontmatter
\else
  \maketitle
  \begin{abstract}
    \paperabstract
  \end{abstract}

  \noindent\textbf{Keywords:} \paperkeywords

\fi

\section{Introduction}

High-fidelity CZ gates are essential to scalable superconducting quantum
processors~\cite{Barends2014,Yan2018,Stehlik2021}. Tunable buses and couplers
provide switchable two-qubit interactions and have enabled high-fidelity CZ
gates~\cite{McKay2016,Mundada2019,Sung2021,Xu2020,Mitchell2021,Marxer2023,Ding2023,Ma2024}.
In implementations based on flux tuning, the conditional phase is generated
by bringing a computational state close to an avoided crossing with a
noncomputational state. Both the accumulated phase and leakage are therefore
sensitive to the flux pulse delivered to the chip.~\cite{Rol2019,Sung2021,Negirneac2021}. Before reaching the chip, the flux pulse is distorted by the finite
bandwidth, impedance mismatches, and long settling tails of the
room-temperature electronics and cryogenic control line.~\cite{Krinner2019,Gustavsson2013,Foxen2019,Rol2020,Guo2024,Hellings2025}.
The flux line therefore has an internal dynamical state set by preceding
controls, making the delivered flux history dependent.~\cite{Rol2019,Gao2021}
We therefore model it as a classical dynamical actuator with memory rather
than a memoryless input-output map.~\cite{Hincks2015}

\begin{figure*}[!t]
  \centering
  \includegraphics[width=\textwidth]{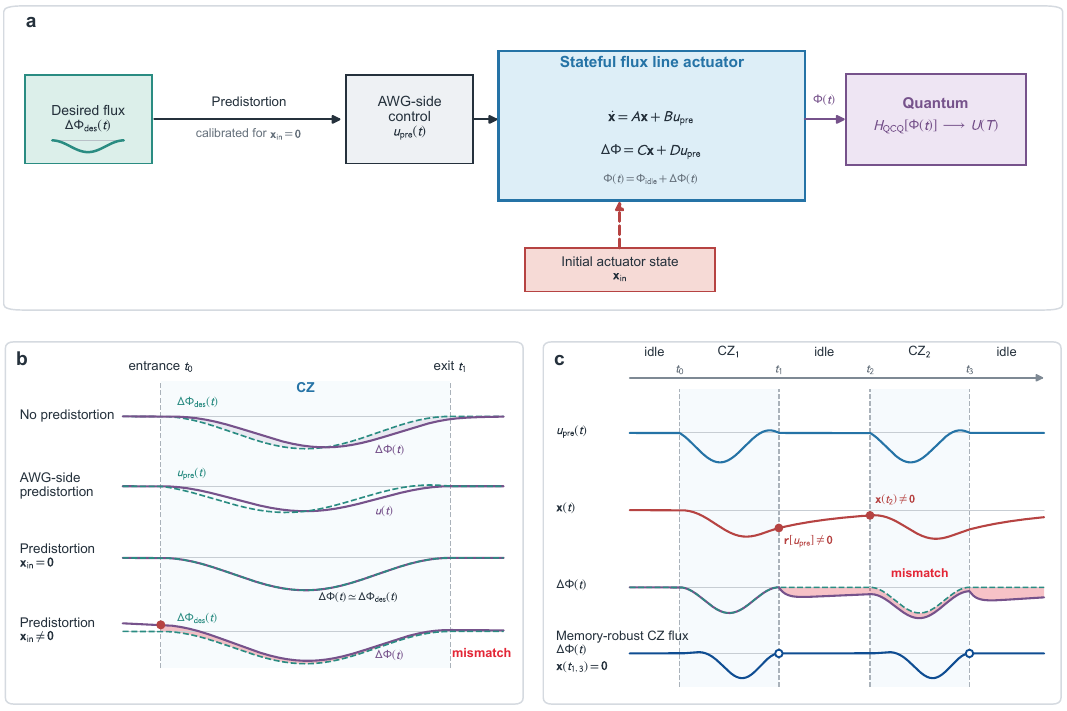}
  \caption{Stateful actuator--quantum dynamics of flux line distortion.
 (a) Predistortion calibrated at the zero initial state,
  $\bm{x}_{\mathrm{in}}=\bm{0}$, maps the desired dynamic flux
  $\Delta\Phi_{\mathrm{des}}(t)$ to the AWG-side input $u_{\rm pre}(t)$. The
  input and initial actuator state jointly determine the delivered dynamic flux
  $\Delta\Phi(t)$ and hence the total flux $\Phi(t)$ that drives the quantum
  system. (b) Without predistortion, the flux line distorts the desired
  waveform. Gate-local predistortion reproduces the desired flux for
  $\bm{x}_{\mathrm{in}}=\bm{0}$, but a nonzero initial state adds the free
  response of the flux line and produces a flux mismatch. (c) In a CZ
  sequence, the actuator residual left at one gate exit evolves through the
  idle interval and changes the flux delivered to the next gate. Memory
  robustness suppresses the gate's first-order sensitivity to the inherited
  initial state, while state closing suppresses the residual written at the
  gate exit, allowing the same predistorted pulse to be repeated. Red shading
  marks the flux mismatch; all waveforms are schematic.}
  \label{fig:stateful-actuator-framework}
\end{figure*}

Measurements with the qubit itself or an embedded cryogenic transducer can
reconstruct the pulse at or near the chip and determine the control line
response~\cite{Gustavsson2013,Foxen2019,Rol2020}. Digital predistortion uses
this characterization to reconstruct the desired flux pulse, while further
calibration against experimental gate errors can reduce the remaining control
errors~\cite{Rol2020,Li2025,Hellings2025,Kelly2014,Egger2014}. Pulse shaping
addresses complementary errors: derivative corrections suppress leakage,
whereas Net-Zero symmetry reduces sensitivity to slow distortions in the flux
line~\cite{Motzoi2009,Rol2019,Negirneac2021}. For conventional gate-local
predistortion, however, calibration normally assumes that the flux line starts
from its zero state. If the line has not fully reset before the next gate,
its nonzero state produces an additional transient, and the calibrated input
no longer reproduces the desired flux pulse. Moreover, the predistortion
calibration does not constrain the flux line state at the end of the pulse,
which becomes the initial state of the next gate.

Addressing these limitations requires a control formulation that explicitly
accounts for the actuator state at the gate boundaries. Robust quantum
control suppresses the effects of Hamiltonian perturbations through numerical
pulse optimization or analytic constructions that cancel leading error
terms~\cite{Khaneja2005,Song2022,Brown2004,Khodjasteh2009,Zeng2019,Buterakos2021,Hai2025,Green2013,Soare2014,Le2022,Propson2022,Poggi2024}.
The response of the control hardware can also be included directly in pulse
optimization~\cite{Hincks2015,Rasulov2025}. Flux line memory, however, is not
represented by a static or quasistatic perturbation in the quantum Hamiltonian.
It instead enters through the initial state of a classical dynamical actuator,
and each component of that state produces a distinct flux perturbation over
time. Robust control of a CZ gate must therefore suppress the errors induced by
variations in the actuator state at the gate entrance and constrain the state
remaining at the gate exit.

To impose these two conditions, we describe the flux line in state space and
couple its dynamics to the quantum system, as shown in
Fig.~\ref{fig:stateful-actuator-framework}. Given the AWG input and the actuator
state at the gate entrance, the coupled model determines the delivered flux,
the quantum propagator, and the actuator state at the gate exit. A first-order
Dyson expansion then associates each component of the initial actuator state
with an error generator in the interaction picture. Memory robustness requires
the computational and leakage components of every generator to vanish, whereas state
closing requires the actuator state written by the pulse to vanish at the gate
exit. The former removes the first-order sensitivity of the current gate to
inherited memory, while the latter prevents the pulse from adding a new
residual to the state passed to the next gate.

We first test the two boundary conditions in the one-pole model. The optimized
pulse exhibits a clear first-order robustness plateau: the infidelity of a
single gate remains near its value at $\bm{x}_{\mathrm{in}}=\bm{0}$ up to
$x_{\mathrm{c}}\simeq3.5\times10^{-3}\Phi_0$ and then grows approximately as
the fourth power of the initial state magnitude. By contrast, the baseline
pulse retains the quadratic scaling expected without first-order cancellation
[Fig.~\ref{fig:one-pole-proof}(c)]. For a sequence of 20 gates, the robust pulse
has a cumulative infidelity of $1.2\times10^{-4}$, compared with $0.79$ for the
baseline pulse [Fig.~\ref{fig:one-pole-proof}(d)]. We then test the method in a
three-pole model fitted to a measured flux line step response~\cite{Li2025}.
The optimized pulse reaches $F_{\mathrm{avg}}=99.9984\%$ for a single gate and
suppresses the error generators associated with all three actuator modes
[Fig.~\ref{fig:three-pole-modal-memory}(c)]. With no additional waiting time
between gates, the complete sequence of ten gates reaches
$F_{\mathrm{avg}}=99.97\%$ [Fig.~\ref{fig:notebook-memory-benchmark}(f)]. Under
the same predistortion protocol, this corresponds to a reduction in sequence
infidelity by a factor of approximately $2.3\times10^{3}$ relative to both the
baseline and Net-Zero pulses. Optimizations with only one boundary condition
separate their roles: memory robustness suppresses errors caused by inherited
actuator states, whereas state closing reduces the residual state at the gate exit generated by each
gate [Fig.~\ref{fig:three-pole-condition-ablation}].

The remainder of the paper is organized as follows.
Section~\ref{sec:forward-model} defines the qubit-coupler-qubit(QCQ) system and the stateful
actuator--quantum forward dynamics.
Section~\ref{sec:robust-gate-boundaries} derives a first-order error generator for each component of the initial actuator state, formulates memory robustness and state closing as separate conditions at the gate entrance and exit, and explains their roles in the repeated CZ gates.
Section~\ref{sec:numerical-implementation} presents the numerical optimization and validation of memory-robust CZ pulses in one-pole and three-pole flux line models. Section~\ref{sec:conclusion} summarizes the conclusions and scope.

\section{Stateful actuator--quantum forward dynamics}
\cleansectionlabel{sec:forward-model}

The AWG control pulse is transmitted through the flux line, whose response
determines the flux applied to the QCQ system. Because the line has internal
dynamics, the delivered flux depends on both the AWG input and the actuator
state at the gate entrance. The same input can therefore produce different
flux pulses for different initial actuator states. To model this dependence,
we first specify the QCQ system under flux control and then describe the flux
line as a classical actuator using a state space representation. We couple the two models to track the quantum evolution and actuator
state during a CZ gate duration. The coupled model, summarized in
Fig.~\ref{fig:stateful-actuator-framework}, provides the boundary quantities
used to derive the robustness conditions in
Sec.~\ref{sec:robust-gate-boundaries}.

\subsection{QCQ system under flux control}
\cleansubsectionlabel{sec:qcq-control-coordinates}

\begin{figure*}[!t]
  \centering
  \includegraphics[width=\textwidth]{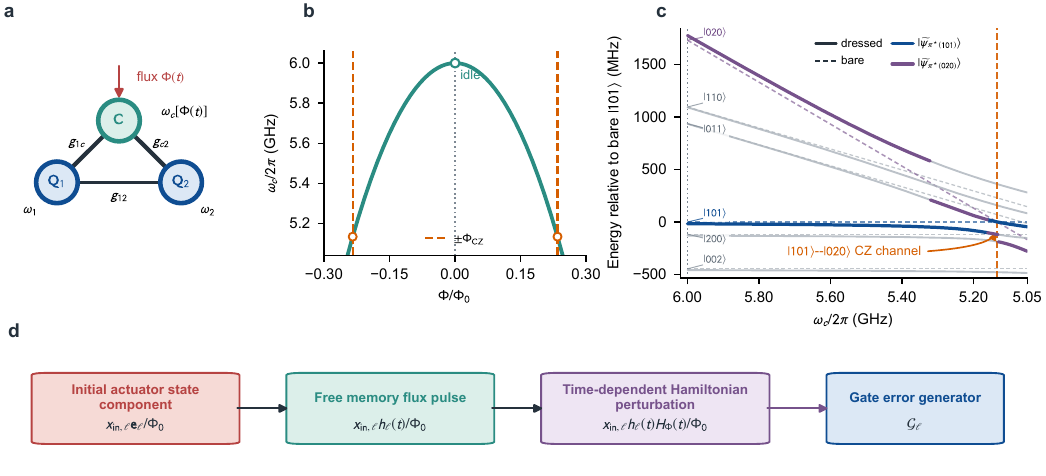}
  \caption{The flux dependence of the tunable coupler determines the CZ
  interaction and the first-order sensitivity of the gate to the initial
  actuator state. (a) Two fixed-frequency qubits coupled
  through a tunable coupler. The delivered flux $\Phi(t)$ tunes the coupler
  frequency $\omega_c[\Phi(t)]$. (b) Flux dependence of $\omega_c/2\pi$ for the
  asymmetric SQUID over the accessible interval $|\Phi|\leq0.30\Phi_0$. The
  gray dotted line marks the idle point, and the orange dashed lines mark
  $\pm\Phi_{\rm CZ}$. (c) Spectrum in the two-excitation manifold as
  $\omega_c/2\pi$ is tuned from its $6.0\,\mathrm{GHz}$ idle value toward the CZ
  interaction. Energies are referenced to the bare $\lvert101\rangle$ level;
  solid and dashed curves denote dressed and bare energies, respectively. Blue
  and purple segments identify
  $\lvert\widetilde{\psi}_{\pi^\star(101)}\rangle$ and
  $\lvert\widetilde{\psi}_{\pi^\star(020)}\rangle$, respectively. The orange
  dashed line marks the resonance of the bare $\lvert101\rangle$ and
  $\lvert020\rangle$ states. (d) A component
  $(x_{\mathrm{in},\ell}/\Phi_0)\bm e_\ell$ of the initial actuator state
  generates the free memory flux
  $(x_{\mathrm{in},\ell}/\Phi_0)h_\ell(t)$ and the resulting time-dependent
  Hamiltonian perturbation
  $(x_{\mathrm{in},\ell}/\Phi_0)h_\ell(t)H_\Phi(t)$. The first-order Dyson
  integral of this perturbation defines the gate error generator
  $\mathcal G_\ell$.}
  \label{fig:qcq-control-sensitivity}
\end{figure*}

The QCQ system consists of two transmon qubits coupled through a tunable coupler. Treating the three transmons as weakly anharmonic oscillators, we write the Hamiltonian as
\begin{align}
H_{\QCQ}(\Phi)
={}&\sum_{j\in\{1,2\}}
\left[
\omega_j n_j+\frac{\alpha_j}{2}n_j(n_j-1)
\right]
\nonumber\\
&+\omega_c(\Phi)n_c+\frac{\alpha_c}{2}n_c(n_c-1)
\nonumber\\
&+\sum_{(j,k)\in\{(1,c),(c,2),(1,2)\}}
g_{jk}\left(a_j^\dagger a_k+a_j a_k^\dagger\right),
\label{eq:qcq-hamiltonian}
\end{align}
where $a_j$ and $n_j=a_j^\dagger a_j$ are the annihilation and number operators
of qubit $j$, respectively.  The qubit frequencies \(\omega_1\) and \(\omega_2\) are fixed, while the coupler frequency \(\omega_c(\Phi)\) depends on the applied flux~\cite{Yan2018,Sung2021}. The parameters \(\alpha_\mu<0\) denote the qubit anharmonicities, and \(g_{jk}\) are the fixed transverse coupling strengths. The total flux \(\Phi(t)\) delivered by the line therefore controls the QCQ dynamics through \(\omega_c[\Phi(t)]\), as illustrated in Fig.~\ref{fig:qcq-control-sensitivity}(a).

Using units with $\hbar=1$, we model the flux dependence of an asymmetric SQUID
coupler as
\begin{equation}
\begin{aligned}
E_{J,c}(\Phi)
&=E_{J\Sigma}
\sqrt{\cos^2\!\left(\pi\Phi/\Phi_0\right)
+d_c^2\sin^2\!\left(\pi\Phi/\Phi_0\right)},\\
\omega_c(\Phi)
&\simeq \sqrt{8E_{C,c}E_{J,c}(\Phi)}-E_{C,c}.
\end{aligned}
\label{eq:coupler-flux-dispersion}
\end{equation}
Here $E_{J\Sigma}=E_{J,L}+E_{J,R}$ is the total Josephson energy,
$d_c=\abs{E_{J,L}-E_{J,R}}/E_{J\Sigma}$ is the junction asymmetry,
$E_{C,c}$ is the coupler charging energy, and $\Phi_0$ is the magnetic flux
quantum~\cite{Koch2007,Hutchings2017}.  An experimentally measured
coupler dispersion $\omega_c(\Phi)$ can replace this analytic form without
changing the coupled model introduced below. For the parameters used here,
Fig.~\ref{fig:qcq-control-sensitivity}(b) shows the nonlinear dependence of
$\omega_c$ on $\Phi$ and the flux values corresponding to the CZ interaction
region.

Because the static couplings remain nonzero at idle point, we define the computational
basis using the four eigenstates of $H_{\QCQ}(\Phi_{\mathrm{idle}})$ that are
adiabatically connected to the corresponding bare states. With ket ordering $(q_1,c,q_2)$, these dressed computational
states are
$\{\lvert\widetilde{000}\rangle,\lvert\widetilde{001}\rangle,
\lvert\widetilde{100}\rangle,\lvert\widetilde{101}\rangle\}$. The projector
$P$ defines the computational subspace $\cH_P=\operatorname{Ran}(P)$, while
$Q=\id-P$ projects onto the orthogonal leakage subspace. Both projectors are
defined at the idle point and held fixed throughout the time evolution and
gate evaluation.

The spectrum of the QCQ system in the two-excitation subspace is shown in
Fig.~\ref{fig:qcq-control-sensitivity}(c). At each $\omega_c$, we label the six
dressed eigenstates using the one-to-one assignment that maximizes the sum of
their squared overlaps with the six bare states
\[
\pi^\star(\omega_c)
=
\arg\max_{\pi}\sum_{\nu}
\bigl|\langle\nu|\widetilde{\psi}_{\pi(\nu)}(\omega_c)\rangle\bigr|^2,
\]
where $\nu$ labels a bare state in this subspace and $\pi$ is a permutation of
the six dressed eigenstates. The blue and purple segments in
Fig.~\ref{fig:qcq-control-sensitivity}(c) identify the eigenstates assigned the
labels $\lvert101\rangle$ and $\lvert020\rangle$, respectively. The assignment
is repeated independently at each $\omega_c$. The $\lvert101\rangle$ and
$\lvert020\rangle$ labels can therefore switch between the two dressed
eigenstates across an avoided crossing, even though their eigenenergies remain
continuous. Within the flux range used for gate
design, the avoided crossing between the states associated with
$\lvert101\rangle$ and $\lvert020\rangle$ produces the conditional energy shift
used to accumulate the conditional phase.

Equations~\eqref{eq:qcq-hamiltonian} and
\eqref{eq:coupler-flux-dispersion} specify how the total flux applied to the
coupler enters the QCQ Hamiltonian, but the AWG sets the input pulse rather
than the flux delivered to the coupler. As shown in
Fig.~\ref{fig:stateful-actuator-framework}(a),
gate-local predistortion determines the AWG-side input $u_{\rm pre}(t)$ needed
to reproduce the desired flux $\Delta\Phi_{\mathrm{des}}(t)$.
The flux line transforms the current $u_{\rm pre}(t)$ into the delivered dynamic flux
$\Delta\Phi(t)$. For a nonzero flux-line state, this output also depends on
earlier controls. The resulting total flux
$\Phi(t)=\Phi_{\mathrm{idle}}+\Delta\Phi(t)$, rather than the desired flux
pulse, drives the QCQ Hamiltonian. Thus, the AWG-side input and the flux-line
state jointly determine the QCQ dynamics as seen in Eq.~\eqref{eq:driven-memory-decomposition}.

\subsection{Coupled actuator and quantum system}

For the bandwidth and signal amplitudes relevant to CZ control, we represent
the flux line as a stable, finite-order, continuous-time linear system in
state-space form~\cite{AstromMurray2008,Ljung1999}. Combining the
actuator state and output equations with the quantum propagator $U(t)$ gives
\begin{subequations}
\label{eq:stateful-actuator-quantum-forward}
\begin{align}
\dot{\bm{x}}(t)
&=A\bm{x}(t)+Bu(t),
\qquad \bm{x}(0)=\bm{x}_{\mathrm{in}},
\label{eq:classical-state}\\
\Delta\Phi(t)
&=C\bm{x}(t)+Du(t),
\label{eq:classical-output}\\
\ii\dot U(t)
&=H_{\QCQ}[\Phi(t)]U(t),
\qquad U(0)=\id.
\label{eq:schrodinger}
\end{align}
\end{subequations}
We refer to Eq.~\eqref{eq:stateful-actuator-quantum-forward} as the
\emph{stateful actuator--quantum forward dynamics}. The coupling is one way:
the actuator determines the flux applied to the QCQ system, while backaction of
the quantum system on the flux line is neglected.

The internal state $\bm{x}(t)\in\mathbb{R}^{n_x}$ carries the memory of earlier
controls. We choose $\bm{x}=\bm{0}$ to represent the fully reset idle state,
so $\bm{x}_{\mathrm{in}}=\bm{x}(0)$ is the actuator state at the gate entrance.
The AWG input $u(t)$ and delivered dynamic flux $\Delta\Phi(t)$ are scalar,
with
$A\in\mathbb{R}^{n_x\times n_x}$,
$B\in\mathbb{R}^{n_x\times 1}$,
$C\in\mathbb{R}^{1\times n_x}$, and $D\in\mathbb{R}$. In the identified
realization, $A$ governs the internal dynamics, $B$ couples the AWG input
into the state dynamics, $C$ maps the state to the delivered flux, and $D$
describes direct feedthrough. The calibrated static offset is included in
$\Phi_{\mathrm{idle}}$, and the model order $n_x$ is chosen to reproduce the
dynamics resolved by end-to-end measurements. We assume that the identified
realization remains time invariant during a gate sequence; slower drift
requires recalibration and is not a noise source in this work.

Solving the linear state-space equations gives an exact decomposition of the state
and output into contributions from the initial actuator state and the applied
input
\begin{equation}
\begin{aligned}
\bm{x}(t)
&=\e^{At}\bm{x}_{\mathrm{in}}
+\int_0^t\e^{A(t-t')}B u(t')\,\dd t',\\
\Delta\Phi(t)
&=C\e^{At}\bm{x}_{\mathrm{in}}
+C\int_0^t\e^{A(t-t')}B u(t')\,\dd t'\\
&\quad+D u(t)\\
&=\Delta\Phi_{\mathrm{memory}}(t)
+\Delta\Phi_{\mathrm{driven}}(t).
\end{aligned}
\label{eq:driven-memory-decomposition}
\end{equation}
Here $\Delta\Phi_{\mathrm{memory}}(t)=C\e^{At}\bm{x}_{\mathrm{in}}$ is the free
memory response, whereas $\Delta\Phi_{\mathrm{driven}}(t)$ contains the
convolution and direct feedthrough terms. This decomposition makes the
limitation of gate-local predistortion explicit. It is
calibrated to reproduce the desired flux through
$\Delta\Phi_{\mathrm{driven}}$ for $\bm{x}_{\mathrm{in}}=\bm{0}$, but does not
compensate for the free response produced by a nonzero initial
state. Pulse reconstruction alone also does
not constrain $\bm{x}(T;\bm{x}_{\mathrm{in}}=\bm{0})$, the actuator state
written at the gate exit. Thus, a pulse calibrated for an isolated gate can
fail when reused in a CZ sequence.

\section{Conditions for CZ gates robust to flux line memory}
\cleansectionlabel{sec:robust-gate-boundaries}

Within the coupled model of
Eq.~\eqref{eq:stateful-actuator-quantum-forward}, flux line memory affects a CZ
sequence at both gate boundaries. The initial actuator state changes the flux
delivered during the current gate, while the driven residual at the gate exit
is inherited by later gates. We first derive an error generator for each
component of the initial state and use these generators to formulate memory
robustness. We then formulate an independent state closing condition for the
driven residual. Finally, a propagation map shows how the two conditions affect
a sequence of CZ gates.

\subsection{From initial actuator state perturbations to quantum error generators}
\cleansubsectionlabel{sec:classical-flux-sensitivities}

Memory robustness is imposed on the quantum gate, whereas
$\bm{x}_{\mathrm{in}}$ describes the initial state of the classical actuator. To connect them, we
propagate a small perturbation of the initial actuator state through the
delivered flux and into the quantum evolution. For a fixed AWG input,
Eq.~\eqref{eq:driven-memory-decomposition} shows that all dependence on
$\bm{x}_{\mathrm{in}}$ enters through the free response. We expand $\bm{x}_{\mathrm{in}}$ in a basis
$\{\bm e_\ell\}_{\ell=1}^{n_x}$. Each basis vector is scaled to a reference
amplitude $\Phi_0$, so that
$\bm{x}_{\mathrm{in}}=\sum_{\ell=1}^{n_x}
(x_{\mathrm{in},\ell}/\Phi_0)\bm e_\ell$ and $x_{\mathrm{in},\ell}$ has units of
flux. In the modal realizations used below, $x_{\mathrm{in},\ell}$ is the
initial stored flux of the $\ell$th state component.

This basis decomposition separates the reference flux trajectory from the flux
perturbation associated with each component of the initial actuator state. Let
$\bar\Phi(t)\equiv\Phi(t;\bm{x}_{\mathrm{in}}=\bm{0})
 =\Phi_{\mathrm{idle}}+\Delta\Phi_{\mathrm{driven}}(t)$ be the reference total
flux produced by the same AWG input for $\bm{x}_{\mathrm{in}}=\bm{0}$.
Equation~\eqref{eq:driven-memory-decomposition} then gives the exact
decomposition
$\Phi(t;\bm{x}_{\mathrm{in}})-\bar\Phi(t)=\sum_{\ell=1}^{n_x}
(x_{\mathrm{in},\ell}/\Phi_0)h_\ell(t)$, where
\begin{equation}
h_\ell(t)
\equiv\left.\Phi_0\frac{\partial\Phi(t;\bm{x}_{\mathrm{in}})}{\partial x_{\mathrm{in},\ell}}\right|_{\bm{x}_{\mathrm{in}}=\bm 0}
=C\e^{At}\bm e_\ell.
\label{eq:initial-state-flux-sensitivity}
\end{equation}
Equation~\eqref{eq:initial-state-flux-sensitivity} defines $h_\ell(t)$ as the
free flux response to the basis displacement $\bm e_\ell$. It has flux units,
and the initial coordinate $x_{\mathrm{in},\ell}$ contributes the physical flux
perturbation $x_{\mathrm{in},\ell}/\Phi_0h_\ell(t)$.

This flux perturbation changes the quantum evolution. Let
$U_0(t)$ be the reference propagator under $\bar\Phi(t)$, and define
$H_\Phi(t)=\left.\partial_\Phi H_{\QCQ}\right|_{\bar\Phi(t)}$ as the
instantaneous Hamiltonian sensitivity to flux along the reference flux trajectory.
A small flux perturbation $\delta\Phi(t)$ produces
$\delta H(t)=\delta\Phi(t)H_\Phi(t)+\cO(\delta\Phi^2)$. An initial state
displacement therefore produces the
time-dependent Hamiltonian perturbation
$(x_{\mathrm{in},\ell}/\Phi_0)h_\ell(t)H_\Phi(t)$. At first order, the Dyson
expansion in the interaction picture gives~\cite{Dyson1949}

\begin{subequations}
\label{eq:initial-state-error-map}
\begin{align}
U(T;x_{\mathrm{in},\ell})
&=U_0(T)\left[\id-\ii\frac{x_{\mathrm{in},\ell}}{\Phi_0}
\mathcal G_\ell\right]
  +\cO\!\left[\left(\frac{x_{\mathrm{in},\ell}}{\Phi_0}\right)^2\right],
\label{eq:terminal-error-generator}\\
\mathcal G_\ell
&=\int_0^T h_\ell(t)
U_0^\dagger(t)H_\Phi(t)U_0(t)\,\dd t .
\label{eq:flux-to-quantum-generator}
\end{align}
\end{subequations}

Here $\mathcal G_\ell$ is the first-order gate error generator in
$x_{\mathrm{in},\ell}/\Phi_0$. The factors $h_\ell(t)$ and $H_\Phi(t)$ describe,
respectively, the free flux response of the actuator and the sensitivity of the
Hamiltonian to flux along the reference flux trajectory. The mapping from the initial
actuator state perturbation to $\mathcal G_\ell$ is shown in
Fig.~\ref{fig:qcq-control-sensitivity}(d).

Composite pulse and geometric control methods commonly treat systematic errors
as static or quasistatic perturbations, although geometric approaches can also
address driven noise~\cite{Brown2004,Zeng2019,Buterakos2021,Hai2025}. Flux line
memory instead generates the structured time-dependent perturbation
$h_\ell(t)H_\Phi(t)$. Equations~\eqref{eq:initial-state-error-map} map this
perturbation to the first-order gate error generator $\mathcal G_\ell$.
Appendix~\ref{sec:supp-propagator-derivative} derives these expressions using
the Dyson expansion.

\subsection{Memory robustness and state closing}
\cleansubsectionlabel{sec:gate-boundary-conditions}

Memory robustness suppresses the first-order logical and leakage errors induced
by each component of a small initial actuator state. State closing separately
suppresses the actuator state written by the pulse at the gate exit.

\subsubsection{Gate entrance: memory robustness}

The first-order expansion in Eq.~\eqref{eq:initial-state-error-map} is linear in
the initial actuator state. For a general small $\bm{x}_{\mathrm{in}}$, the
total first-order error generator is
$\sum_{\ell=1}^{n_x}(x_{\mathrm{in},\ell}/\Phi_0)\mathcal G_\ell$.
Because the components $x_{\mathrm{in},\ell}$ vary independently, robustness
requires the relevant blocks of every $\mathcal G_\ell$ to vanish separately.
For a reference evolution that realizes the target CZ gate and preserves
$\cH_P$, these are the traceless logical block and the leakage block. With
$\dim(\cH_P)=4$, we therefore impose
\begin{equation}
\begin{aligned}
\mathcal G_\ell^{\mathrm{log}}
&\equiv P\mathcal G_\ell P
-\frac{\Tr(P\mathcal G_\ell P)}{4}P=0,\\
\mathcal G_\ell^{\mathrm{leak}}
&\equiv Q\mathcal G_\ell P=0,
\qquad \ell=1,\ldots,n_x.
\end{aligned}
\label{eq:memory-transparency-conditions}
\end{equation}
The first condition eliminates the traceless logical error generator, leaving
only an irrelevant global phase, whereas the second eliminates first-order
coupling from the computational subspace to leakage subspace. Equation~\eqref{eq:memory-transparency-conditions}
is imposed for every actuator state component through its free response
$h_\ell(t)$. We use the $\Phi_0$-normalized basis defined above. These
conditions constrain the first-order error accumulated during the present gate and
they do not constrain the actuator state left at its exit.

\subsubsection{Gate exit: state closing}

The gate exit condition follows directly from the actuator solution.
Evaluating Eq.~\eqref{eq:driven-memory-decomposition} at $t=T$ gives the exact
map from the initial state to the terminal state,
\begin{equation}
\begin{aligned}
\bm{x}(T;\bm{x}_{\mathrm{in}})
&=\e^{AT}\bm{x}_{\mathrm{in}}+\bm r[u],\\
\bm r[u]
&=\int_0^T\e^{A(T-t')}Bu(t')\,\dd t'.
\end{aligned}
\label{eq:residual-state}
\end{equation}
Equation~\eqref{eq:residual-state} separates the inherited and driven
contributions to the terminal state. The first term propagates the initial
actuator state, whereas $\bm r[u]$ is the state written by the present AWG
input. When $\bm{x}_{\mathrm{in}}=\bm{0}$, the terminal state is $\bm r[u]$,
so state closing requires
\begin{equation}
\bm{x}(T;\bm{x}_{\mathrm{in}}=\bm{0})
=\bm r[u]=\bm{0}.
\label{eq:robust-state-closing}
\end{equation}
State closing prevents the AWG input from writing a residual actuator state,
but it does not remove an arbitrary initial state. 

\subsection{Propagation of flux line memory in repeated CZ gates}
\cleansubsectionlabel{sec:sequence-memory-scope}

In a sequence of consecutive CZ gates, the actuator state at the exit of one
gate becomes the initial state of the next. Let $\bm{x}^{(n)}$ denote the state
after $n$ completed gates, and hence at the entrance to gate $n+1$, with
$\bm{x}^{(0)}=\bm{x}_{\mathrm{in}}$. Defining the state transition over one
gate as $A_T\equiv\e^{AT}$, reuse of the same AWG input in
Eq.~\eqref{eq:residual-state} gives
\begin{equation}
\begin{aligned}
\bm{x}^{(n+1)}
&=A_T\bm{x}^{(n)}+\bm r[u],\\
\bm{x}^{(n)}
&=A_T^n\bm{x}_{\mathrm{in}}
+\sum_{q=0}^{n-1}A_T^q\bm r[u] .
\end{aligned}
\label{eq:discrete-state-map}
\end{equation}
The recurrence separates two contributions to the actuator state. The first
term propagates the state present before the sequence through $n$ gates, whereas
the sum accumulates the driven residual $\bm r[u]$ written by each gate.

When state closing holds, the driven residual does not
accumulate, although the inherited state $A_T^n\bm{x}_{\mathrm{in}}$ can remain
nonzero. Memory robustness suppresses the first-order gate error induced by
each component of this inherited state, but it does not alter
Eq.~\eqref{eq:discrete-state-map}. Without state closing, a nonzero $\bm r[u]$
can accumulate and move the actuator state beyond the range in which the
first-order expansion is accurate. Both conditions are therefore required to
reuse the same pulse across a CZ gates sequence. Figure~\ref{fig:three-pole-condition-ablation}
tests their distinct roles using pulses optimized for only one condition.

\section{Numerical validation of memory-robust CZ gates}
\cleansectionlabel{sec:numerical-implementation}

  We optimize finite-basis CZ pulses under the memory-robustness and state-closing conditions of Sec.~\ref{sec:robust-gate-boundaries}. A one-pole model tests first-order cancellation in the simplest stateful line, and a three-pole model fitted to an experimental step response introduces multiple memory time scales. Finite initial-state scans, continuously propagated CZ sequences, and single-condition optimizations are reserved for validation and do not enter the pulse objective.

\subsection{Numerical setup and pulse optimization}
\cleansubsectionlabel{sec:propagation-validation}

The numerical simulations use the Hamiltonian in
Eq.~\eqref{eq:qcq-hamiltonian} with representative parameters in the range
reported for tunable coupler devices~\cite{Yan2018,Sung2021}:
\begin{align*}
 (\omega_1,\omega_2)/2\pi&=(5.064,4.904)~\mathrm{GHz},\\
 \omega_c(\Phi_{\mathrm{idle}})/2\pi&=6.000~\mathrm{GHz},\\
 (\alpha_1,\alpha_c,\alpha_2)/2\pi
   &=(-0.276,-0.300,-0.286)~\mathrm{GHz},\\
 (g_{1c},g_{2c},g_{12})/2\pi
   &=(0.090,0.090,0.00796)~\mathrm{GHz},
\end{align*}
The SQUID asymmetry is $d_c=0.10$. Each transmon is truncated to three local
levels. Because Eq.~\eqref{eq:qcq-hamiltonian} conserves the total excitation
number $N=n_1+n_c+n_2$, numerical propagation uses the ten-dimensional
$N\leq2$ invariant subspace. The gate duration is $T=50~\mathrm{ns}$
throughout.

We parameterize the desired flux pulse as
\begin{equation}
 \Delta\Phi_{\mathrm{des}}(t)
 =\Phi_{\max}\tanh\!\left[
 \sum_{m=1}^{M}c_m\sin(m\pi t/T)
 \right],
\label{eq:notebook-control-parameterization}
\end{equation}
where $M$ is the number of sine basis functions and
$\bm c=(c_1,\ldots,c_M)^{\mathsf T}$ contains the dimensionless pulse
coefficients. This parameterization gives
$\Delta\Phi_{\mathrm{des}}(0)=\Delta\Phi_{\mathrm{des}}(T)=0$ and enforces
$\abs{\Delta\Phi_{\mathrm{des}}(t)}<\Phi_{\max}$, with
$\Phi_{\max}=0.30\Phi_0$. For each calibrated flux line model, gate-local
predistortion maps $\Delta\Phi_{\mathrm{des}}(t)$ to $u_{\rm pre}(t)$ so that
$\Delta\Phi(t)=\Delta\Phi_{\mathrm{des}}(t)$ when
$\bm{x}_{\mathrm{in}}=\bm 0$. A nonzero initial actuator state adds the free
memory response, so the same predistorted input no longer reproduces the
desired flux pulse.

We calculate the average gate fidelity, including leakage, as
\begin{equation}
F_{\mathrm{avg}}
=\frac{\Tr(K^\dagger K)
+\abs{\Tr(U_{\mathrm{tar}}^\dagger K)}^2}{20}.
\label{eq:numerical-gate-metrics}
\end{equation}
Here $K=P U(T)P|_{\cH_P}$ is the propagator restricted to the four-dimensional
computational subspace.
A virtual-$Z$ correction is calibrated once at $\bm{x}_{\mathrm{in}}=\bm 0$
and incorporated into $U_{\mathrm{tar}}$. For a sequence of $n$ CZ gates, we
propagate the coupled flux line and quantum dynamics continuously from $t=0$
to $nT$, with no waiting interval between gates. We then evaluate
Eq.~\eqref{eq:numerical-gate-metrics} with $K=P U(nT)P|_{\cH_P}$ and target
$U_{\mathrm{tar}}^n$; the resulting cumulative infidelity $1-F_{\mathrm{avg}}$
is plotted for the one-pole model in Fig.~\ref{fig:one-pole-proof}(d).

To impose the memory robustness and state closing conditions in
Eqs.~\eqref{eq:memory-transparency-conditions} and
\eqref{eq:robust-state-closing}, we minimize
\begin{equation}
\begin{aligned}
 \mathcal J(\bm c)
 &=w_F\bigl[1-F_{\mathrm{avg}}(\bm c)\bigr]
   +\sum_{\ell=1}^{n_x}w_{\ell}^{\mathrm{log}}
   \norm{\mathcal G_{\ell}^{\mathrm{log}}}_{\mathrm F}^{2}
   \\
 &\quad+\sum_{\ell=1}^{n_x}w_{\ell}^{\mathrm{leak}}
   \norm{\mathcal G_{\ell}^{\mathrm{leak}}}_{\mathrm F}^{2}
   \\
 &\quad+\sum_{\ell=1}^{n_x}w_{\ell}^{r}
   \abs{\frac{r_\ell[u]}{\Phi_0}}^2.
\end{aligned}
\label{eq:common-pulse-objective}
\end{equation}
Here $w_F$, $w_\ell^{\mathrm{log}}$, $w_\ell^{\mathrm{leak}}$, and $w_\ell^r$
are the relative weights of the fidelity, logical generator, leakage
generator, and state closing terms, respectively. The quantity $r_\ell[u]$ is
the $\ell$th component of $\bm r[u]$ in Eq.~\eqref{eq:residual-state}. The
generator norms are dimensionless, and each residual component is normalized
explicitly by $\Phi_0$. The weights absorb
the relative numerical scales of these terms. The fidelity term is evaluated
at $\bm{x}_{\mathrm{in}}=\bm 0$. We use the bound-constrained L-BFGS-B
algorithm~\cite{Morales2011,Virtanen2020}. The weights and pulse initialization
are given in Appendix~\ref{sec:appendix-optimization}.

\subsection{One-pole demonstration of memory robustness}
\cleansubsectionlabel{sec:one-pole-proof}

\begin{figure*}[!t]
  \centering
  \includegraphics[width=0.9\textwidth]{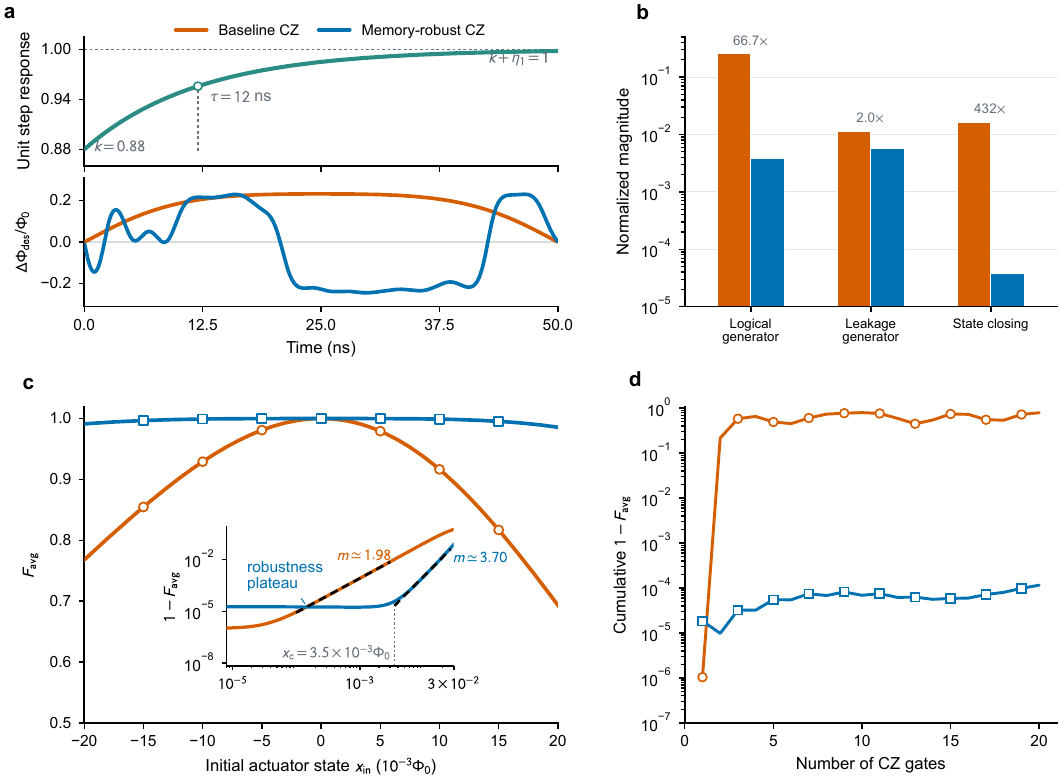}
  \caption{One-pole validation of memory robustness and state closing.
  (a) Unit step response of the one-pole flux line model (upper) and desired flux pulses
  optimized for the baseline and memory-robust CZ gates (lower).
  (b) Normalized magnitudes of the logical and leakage error generators and the
  state closing residual. Labels above the bars give the reduction
  factors relative to the baseline.
  (c) Average fidelity versus the initial actuator state.
  The inset shows $1-F_{\rm avg}$ for $x_{\rm in}>0$ on logarithmic axes.
  $x_{\rm c}$ marks the end of the first-order robustness plateau, and the
  dashed lines are power law fits to $1-F_{\rm avg}\propto x_{\rm in}^{m}$.
  (d) Cumulative infidelity for 20 consecutive CZ gates with no waiting interval
  between gates. At gate 20, $1-F_{\rm avg}=0.79$ for the baseline pulse and
  $1.2\times10^{-4}$ for the memory-robust pulse.}
  \label{fig:one-pole-proof}
\end{figure*}

The one-pole flux line model provides the simplest setting in which to test
memory robustness and state closing:
\begin{equation}
 \dot{x}_1=-\frac{x_1}{\tau_1}+\frac{\eta_1}{\tau_1}u,
 \qquad
 \Delta\Phi=ku+x_1,
 \label{eq:one-pole-proof-line}
\end{equation}
with $(k,\eta_1,\tau_1)=(0.88,0.12,12~\mathrm{ns})$. Because the direct and
pole weights satisfy $k+\eta_1=1$, the model has unit steady-state gain. Its
unit step response is shown in the upper part of
Fig.~\ref{fig:one-pole-proof}(a). Since the model has a single actuator state
component, the objective contains one logical generator norm, one leakage
generator norm, and one state closing term. The baseline pulse is parameterized
by 5 sine basis functions and is optimized only for CZ fidelity at
$\bm{x}_{\mathrm{in}}=\bm 0$. The memory-robust pulse is parameterized by 20
sine basis functions and is optimized with
Eq.~\eqref{eq:common-pulse-objective}. The corresponding desired flux pulses
are shown in the lower part of Fig.~\ref{fig:one-pole-proof}(a).

The memory-robust pulse reduces the logical and leakage generator norms by
factors of 66.7 and 2.0, respectively, and the state closing residual by a
factor of 432 [Fig.~\ref{fig:one-pole-proof}(b)]. The scan over the scalar
initial actuator state shows a first-order robustness plateau up to
$x_{\rm c}\simeq3.5\times10^{-3}\Phi_0$. Beyond $x_{\rm c}$, the infidelity
has approximately quartic scaling with $m\simeq3.70$, whereas the baseline
pulse shows the expected quadratic scaling with $m\simeq1.98$
[Fig.~\ref{fig:one-pole-proof}(c)]. When the same predistorted AWG input is
applied in 20 consecutive gates with no waiting interval, the cumulative
infidelity remains $1.2\times10^{-4}$, compared with $0.79$ for the baseline
pulse [Fig.~\ref{fig:one-pole-proof}(d)]. These results provide direct
numerical validation of the two design conditions derived in
Sec.~\ref{sec:robust-gate-boundaries} for constructing CZ gates that are robust
to flux line memory.

\subsection{Memory-robust CZ gate in a three-pole flux line model}

\begin{figure*}[!t]
  \centering
  \includegraphics[width=0.9\textwidth]{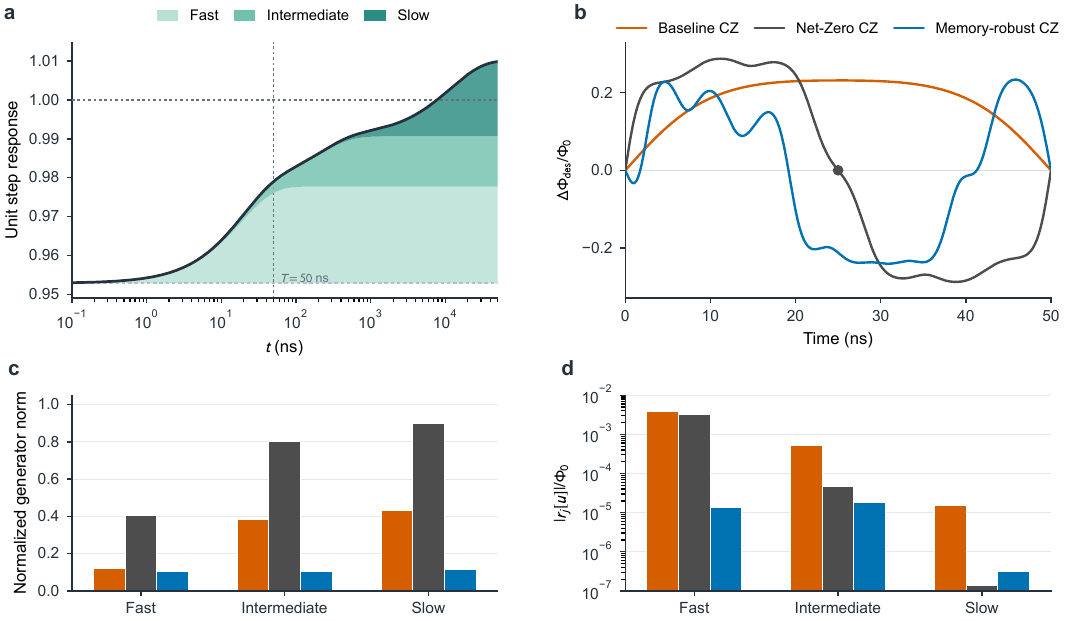}
  \caption{Memory-robust CZ pulse in a
  three-pole flux line model. (a) Unit step response of the fitted model
  (black) and the fast, intermediate, and slow modal contributions (shaded).
  The horizontal dashed lines mark the direct feedthrough $k$ and unity; the
  vertical dashed line marks the gate duration $T=50~\mathrm{ns}$.
  (b) Desired flux pulses for the baseline, Net-Zero, and memory-robust CZ
  gates. The marker at $t=T/2$ identifies the zero crossing imposed by
  Net-Zero antisymmetry. (c) Combined logical and leakage generator norms for
  each actuator mode, evaluated for an initial actuator state displacement
  $x_{\mathrm{in},\ell}=10^{-3}\Phi_0$. (d) Magnitudes of the gate exit
  actuator state components for $\bm{x}_{\mathrm{in}}=\bm 0$, normalized by
  $\Phi_0$.}
  \label{fig:three-pole-modal-memory}
\end{figure*}

To represent multiple memory time scales, we use a three-pole flux line model
fitted to the measured step response reported by Li \emph{et al.}~\cite{Li2025}.
For $\bm{x}=(x_1,x_2,x_3)^{\mathsf T}$, the state space matrices in
Eqs.~\eqref{eq:classical-state} and
\eqref{eq:classical-output} are
\begin{equation*}
\begin{gathered}
A=-\operatorname{diag}\!\left(\frac{1}{\tau_1},
\frac{1}{\tau_2},\frac{1}{\tau_3}\right),\qquad
B=\begin{pmatrix}
 \eta_1/\tau_1\\ \eta_2/\tau_2\\ \eta_3/\tau_3
\end{pmatrix},\\
C=\begin{pmatrix}1&1&1\end{pmatrix},\qquad D=k,
\end{gathered}
\end{equation*}
with fitted parameters
\begin{align*}
 k&=0.9528,\\
 (\eta_1,\eta_2,\eta_3)&=(0.02487,0.01290,0.01955),\\
 (\tau_1,\tau_2,\tau_3)&=(18.47~\mathrm{ns},199.4~\mathrm{ns},
 11.80~\mu\mathrm{s}).
\end{align*}
In this diagonal realization, $x_\ell$ is the stored flux contributed by the
$\ell$th actuator mode. Figure~\ref{fig:three-pole-modal-memory}(a) shows the
unit step response and the three modal contributions. Relative to the gate
duration $T=50~\mathrm{ns}$, the free response of the fast mode largely decays
within one gate, the intermediate response persists over several gates, and the
slow response changes little over the simulated sequence.

Each mode contributes one independent component to the initial actuator state.
Equation~\eqref{eq:common-pulse-objective} therefore includes one logical
generator term, one leakage generator term, and one state closing term for each
mode. We compare the resulting memory-robust pulse with two high-fidelity
references under the same quantum and flux line models. The baseline pulse uses
5 sine basis functions and is also shown in
Fig.~\ref{fig:one-pole-proof}(a). The Net-Zero pulse uses 10 even-index sine
basis functions and satisfies
$\Delta\Phi_{\mathrm{des}}(T-t)=-\Delta\Phi_{\mathrm{des}}(t)$, following the
antisymmetric construction of Ref.~\cite{Rol2019}. Both reference pulses are
optimized only for CZ fidelity at $\bm{x}_{\mathrm{in}}=\bm 0$ and exceed
$99.999\%$. The memory-robust pulse uses 20 sine basis functions and reaches
$F_{\mathrm{avg}}=99.9984\%$. Figure~\ref{fig:three-pole-modal-memory}(b)
compares the three optimized flux pulses.

The memory-robust pulse gives the smallest combined logical and leakage
generator norm for each actuator mode [Fig.~\ref{fig:three-pole-modal-memory}(c)].
Relative to the baseline, the reduction is modest for the fast mode and much
larger for the intermediate and slow modes; the Net-Zero pulse gives no
comparable suppression. The residual components in
Fig.~\ref{fig:three-pole-modal-memory}(d) correspond to a gate exit actuator
state norm of $2.23\times10^{-5}\Phi_0$. Thus, the optimization suppresses both
the sensitivity to the initial actuator state and the gate exit residual across
all three memory time scales within a single 50-ns gate.

\subsection{Benchmarks under initial state perturbations and repeated CZ gates}

The local generator norms and gate exit residual components in
Figs.~\ref{fig:three-pole-modal-memory}(c) and (d) do not determine the range of
initial actuator states over which high fidelity is retained or the performance
of repeated CZ gates. We therefore test the three optimized pulses under finite
perturbations of the initial actuator state and in consecutive gate sequences.
Each sequence reuses the same gate-local predistorted AWG input without an
additional waiting interval for the actuator to reset. As a reference that uses
full state information, the state-tracking inverse reproduces the baseline
desired flux sequence by updating the AWG input using the exact evolving
actuator state.

\ifdefined\TwoColumnWideAbstract
\begin{figure*}[!t]
\else
\begin{figure}[H]
\fi
  \centering
  \includegraphics[width=\textwidth]{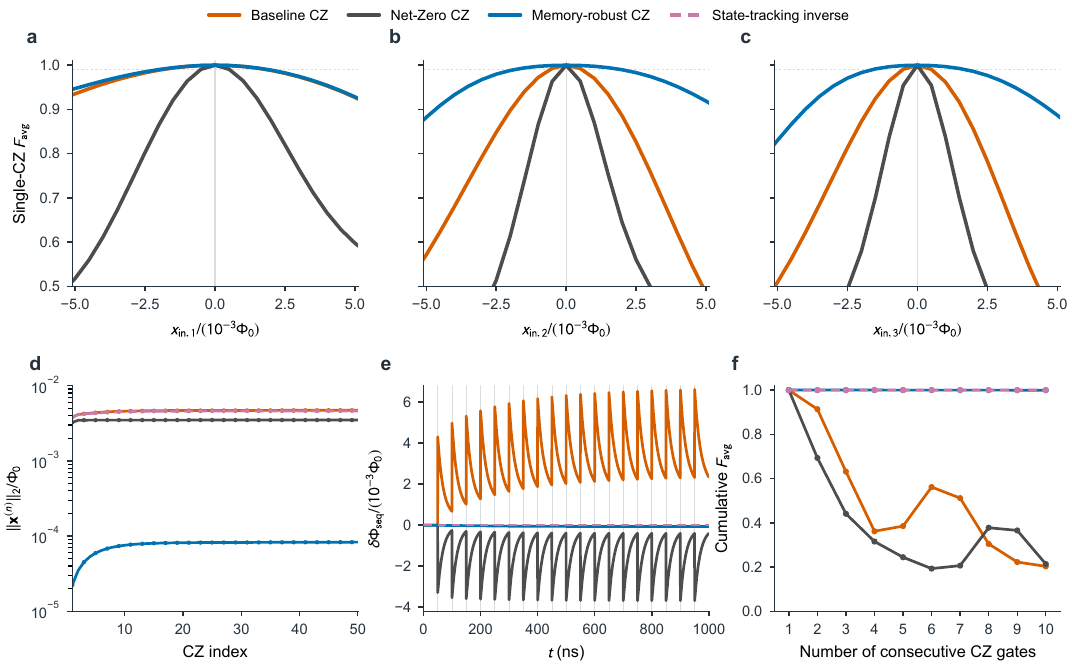}
  \caption{Validation under finite perturbations of the initial actuator state
  and repeated CZ gates. (a)--(c) Average gate fidelity of a
  single CZ gate when each component of $\bm{x}_{\mathrm{in}}$ is varied
  separately, with the other components set to zero.
  (d) Actuator state norm after each of 50 consecutive gates starting from
  $\bm{x}_{\mathrm{in}}=\bm 0$. (e) Sequence flux mismatch
  $\delta\Phi_{\mathrm{seq}}(t)=\Delta\Phi(t)-
  \Delta\Phi_{\mathrm{des,seq}}(t)$ over 20 gates; vertical lines mark the gate
  boundaries. The three fixed pulses reuse their respective predistorted inputs,
  whereas the state-tracking inverse reproduces the baseline target sequence
  using the exact evolving actuator state. (f) Cumulative average gate fidelity
  after each of the first ten gates.}
  \label{fig:notebook-memory-benchmark}
\ifdefined\TwoColumnWideAbstract
\end{figure*}
\else
\end{figure}
\fi

The free response associated with the fast component largely decays during a
50-ns gate, so the memory-robust pulse provides only a modest improvement over
the baseline for perturbations in $x_{\mathrm{in},1}$. The intermediate and slow
responses persist beyond one gate, and the memory-robust pulse retains high
fidelity over much wider ranges of $x_{\mathrm{in},2}$ and
$x_{\mathrm{in},3}$ than either reference pulse
[Figs.~\ref{fig:notebook-memory-benchmark}(a)--(c)]. This behavior agrees with
the generator norms in Fig.~\ref{fig:three-pole-modal-memory}(c).

Figures~\ref{fig:notebook-memory-benchmark}(d)--(f) connect the two gate
boundary conditions to sequence fidelity. Starting from
$\bm{x}_{\mathrm{in}}=\bm 0$, all three fixed inputs reconstruct their first
desired flux pulse to numerical precision. State closing then keeps the
actuator state at the exit of each memory-robust gate below $10^{-4}\Phi_0$ over
50 gates [Fig.~\ref{fig:notebook-memory-benchmark}(d)]. Consequently, its RMS
flux mismatch is 59 times smaller than that of the baseline and remains close to
zero on the scale of Fig.~\ref{fig:notebook-memory-benchmark}(e). The
state-tracking inverse also cancels the mismatch, but does so by updating the
AWG input using the exact evolving state even when the actuator state is
nonzero. The suppressed error generators of the memory-robust pulse further
reduce the quantum error caused by its remaining mismatch. After ten gates, its
cumulative fidelity is $99.97\%$, compared with $20.3\%$ for the baseline and
$21.4\%$ for the Net-Zero pulse
[Fig.~\ref{fig:notebook-memory-benchmark}(f)], corresponding to an approximately
$2.3\times10^3$ reduction in sequence infidelity relative to both references.
Because the reference pulses have higher fidelity for an isolated gate at
$\bm{x}_{\mathrm{in}}=\bm 0$, their sequence deterioration arises from actuator
memory rather than from the calibration of the first gate.

\subsection{Necessity of both gate-boundary conditions}

To isolate the role of each boundary condition, we optimize two auxiliary
pulses that retain either the six generator penalties or the three state closing
penalties, together with the fidelity term. The auxiliary and combined pulses
share the gate duration, flux bound, 20 sine basis functions, gate-local
predistortion, and initial coefficients.

\ifdefined\TwoColumnWideAbstract
\begin{figure*}[!t]
\else
\begin{figure}[H]
\fi
  \centering
  \includegraphics[width=0.9\textwidth]{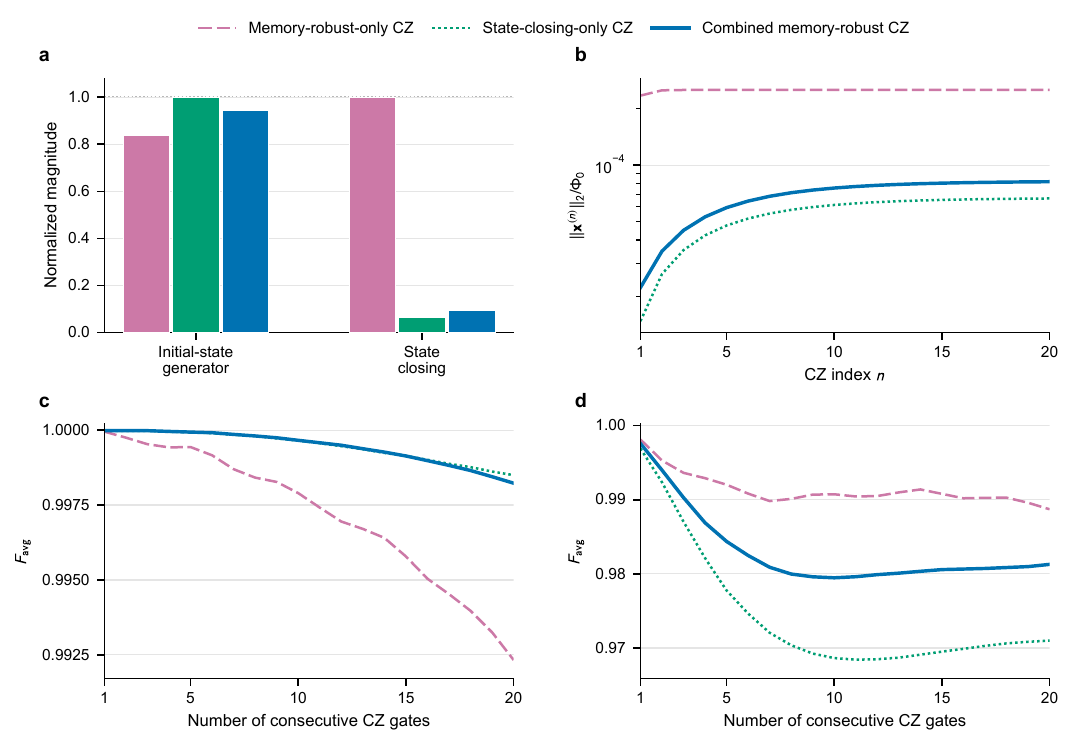}
  \caption{Comparison of pulses optimized for memory robustness, state closing,
  or both in the three-pole model. All pulses use a 50-ns duration, a
  $0.30\Phi_0$ flux bound, 20 sine basis functions, the same initialization,
  and calibrated gate-local predistortion. At
  $\bm{x}_{\mathrm{in}}=\bm 0$, all pulses reach
  $F_{\mathrm{avg}}>99.994\%$. (a) Aggregate initial state generator magnitude,
  normalized to the pulse optimized for state closing alone, and gate exit
  residual norm, normalized to the pulse optimized for memory robustness alone.
  (b) Actuator state norm and (c) cumulative average gate fidelity over 20 gates
  starting from $\bm{x}_{\mathrm{in}}=\bm 0$. (d) Cumulative average gate
  fidelity for $x_{\mathrm{in},2}=10^{-3}\Phi_0$, with the other components set
  to zero. Each sequence reuses its fixed predistorted input without an
  additional waiting interval. The sequence trajectories are excluded from
  optimization.}
  \label{fig:three-pole-condition-ablation}
\ifdefined\TwoColumnWideAbstract
\end{figure*}
\else
\end{figure}
\fi

Figure~\ref{fig:three-pole-condition-ablation}(a) separates the two conditions.
Memory robustness alone minimizes the aggregate generator magnitude but leaves
the largest gate exit residual, whereas state closing alone does the reverse.
The combined pulse suppresses both quantities without reaching either
single-condition minimum.

Panels (b)--(d) show the consequences for a CZ sequence. At
$\bm{x}_{\mathrm{in}}=\bm 0$, the first gate has no inherited actuator state;
the CZ sequence therefore reveals how gate exit residuals generate memory.
The two
pulses that include state closing maintain smaller boundary states and higher
cumulative fidelities than memory robustness alone
[Figs.~\ref{fig:three-pole-condition-ablation}(b) and (c)]. At
$x_{\mathrm{in},2}=10^{-3}\Phi_0$, state closing does not suppress the quantum
error induced by the inherited intermediate component. Memory robustness alone
therefore gives the highest fidelity, while the combined pulse gives an intermediate result
[Fig.~\ref{fig:three-pole-condition-ablation}(d)]. Thus, neither condition alone
controls both terms in Eq.~\eqref{eq:discrete-state-map}. The combined pulse
limits the driven residual while reducing sensitivity to an inherited actuator
state.

\section{Conclusion and outlook}
\cleansectionlabel{sec:conclusion}

We treat flux line memory as part of the coupled dynamics of a classical
actuator and a quantum system. The resulting first-order analysis yields two
complementary requirements at the gate boundaries. The quantum evolution must
be first-order insensitive to each component of the initial actuator state,
while the control pulse must leave a small driven actuator state at the gate
exit. The first condition limits how inherited memory affects the current gate;
the second limits the residual actuator state generated by that gate and passed
to the next.

The one-pole model exhibits the predicted first-order robustness plateau, and
the three-pole model shows that both boundary requirements can be met
simultaneously for a line with several memory time scales. The same
predistorted pulse can then be reused without an additional settling interval:
the complete ten-gate sequence retains an average gate fidelity of $99.97\%$,
and its infidelity is smaller by a factor of about $2.3\times10^{3}$ than those
of the baseline and Net-Zero pulses. The single-condition comparison
further shows that neither generator suppression nor state closing alone
controls both routes by which memory affects the sequence.

Within a calibrated finite-dimensional linear model, these results show that
gate-local predistortion need not be replaced by a full state tracking inverse
over the entire sequence; memory can instead be addressed during gate pulse
optimization. Experimental validation should next determine how model
identification errors, drift, nonlinear line response, flux noise, and
decoherence affect the achieved sequence fidelity. Experiments with variable
gate spacings and interleaved operations would further test whether the
protection persists in less regular schedules. The same formulation may also
apply to other flux control operations whose actuator response persists beyond
the nominal gate window.

\FloatBarrier
\appendix
\numberwithin{equation}{section}
\numberwithin{figure}{section}
\numberwithin{table}{section}
\renewcommand{\thesection}{Appendix~\Alph{section}:}
\renewcommand{\theequation}{\Alph{section}\arabic{equation}}
\renewcommand{\thefigure}{\Alph{section}\arabic{figure}}
\renewcommand{\thetable}{\Alph{section}\arabic{table}}
\renewcommand{\thesubsection}{\Alph{section}.\arabic{subsection}}
\renewcommand{\thesubsubsection}{\thesubsection.\arabic{subsubsection}}

\section{Optimization procedure}
\cleanappendixlabel{sec:appendix-optimization}

\begin{table*}[!t]
\centering
\small
\caption{Loss weights used in the reported pulse optimizations.}
\label{tab:optimization-weights}
\begin{tabular}{lcccc}
\hline
Pulse & $w_F$ & $\bm w^{\rm log}$ & $\bm w^{\rm leak}$ & $\bm w^r$ \\
\hline
One-pole robust & 10 & 20 & 20 & 0.1 \\
Three-pole robust & 5 & $(1,3,5)$ & $(3,5,6)$ & $(2,1,0.5)$ \\
Memory robustness only & 1.256 & $(0.165,0.772,1.823)$ & $(1.238,1.724,2.014)$ & $(0,0,0)$ \\
State closing only & 2.5 & $(0,0,0)$ & $(0,0,0)$ & $(1.167,2.334,0.367)$ \\
\hline
\end{tabular}
\end{table*}

The coefficients $c_m$ in Eq.~\eqref{eq:notebook-control-parameterization} are
optimized by L-BFGS-B with gradients from automatic differentiation and bounds
$-6\leq c_m\leq6$. The baseline search starts from the lowest-order sine basis
function, whereas the Net-Zero search is restricted to even indices to enforce
antisymmetry. The memory-robust searches vary all coefficients and use random
vectors or previously converged pulses as initial guesses.

All terms retained in Eq.~\eqref{eq:common-pulse-objective} are active
throughout each memory-robust search. Finite initial actuator state
perturbations and repeated gate trajectories are used only for validation. The
single-condition searches follow the same procedure with either the generator
or state closing penalties omitted.

Table~\ref{tab:optimization-weights} lists the weights used in the optimization
stages that produced the pulses reported in
\mbox{Figs.~\ref{fig:one-pole-proof}--\ref{fig:three-pole-condition-ablation}}.
For the one-pole search, the generator norms are evaluated at
$x_{\mathrm{in},1}=4\times10^{-3}\Phi_0$, which also normalizes the gate exit
state. The three-pole searches use a
scale-compressed form of Eq.~\eqref{eq:common-pulse-objective}, with fidelity,
generator, and state closing contributions given by
$\log[1+(1-F_{\mathrm{avg}})/\epsilon_F]$,
$\log[1+(10^{-3}\norm{\mathcal G_\ell^s}_{\mathrm F}/g_0)^2]$, and
$\log[1+(r_\ell[u]/r_0)^2]$, respectively. Here
$s\in\{\mathrm{log},\mathrm{leak}\}$, $\epsilon_F=5\times10^{-5}$,
$g_0=0.05$, and $r_0=5\times10^{-5}\Phi_0$. A fidelity barrier and weak
smoothness terms stabilize the search without changing the gate boundary
conditions. Weight vector entries are ordered by the fast, intermediate, and
slow actuator modes.

Table~\ref{tab:robust-pulse-coefficients} gives the dimensionless coefficients
of the primary three-pole memory-robust pulse. Together with
Eq.~\eqref{eq:notebook-control-parameterization}, they define the desired flux pulse in
Figs.~\ref{fig:three-pole-modal-memory}--\ref{fig:three-pole-condition-ablation}.

\begin{table}[ht]
\centering
\small
\caption{Dimensionless coefficients of the three-pole memory-robust CZ pulse.}
\label{tab:robust-pulse-coefficients}
\begin{tabular}{r r r r}
\hline
$m$ & $c_m$ & $m$ & $c_m$ \\
\hline
1  & $-0.36880843$ & 11 & $ 0.01474042$ \\
2  & $ 0.50662224$ & 12 & $-0.01618463$ \\
3  & $ 0.68738861$ & 13 & $ 0.06440639$ \\
4  & $-0.27119782$ & 14 & $ 0.00600673$ \\
5  & $ 0.23005490$ & 15 & $-0.12386559$ \\
6  & $-0.13206655$ & 16 & $-0.06142061$ \\
7  & $ 0.19703598$ & 17 & $-0.08503133$ \\
8  & $ 0.00661433$ & 18 & $-0.02531204$ \\
9  & $ 0.03337576$ & 19 & $-0.01808988$ \\
10 & $-0.12097157$ & 20 & $-0.05498015$ \\
\hline
\end{tabular}
\end{table}

\FloatBarrier
\section{Derivation of the initial-state error generator}
\cleanappendixlabel{sec:supp-propagator-derivative}

Vary one component $x_{\mathrm{in},\ell}$ of the initial actuator state about
zero while keeping all other components fixed at zero. Define the corresponding
derivative of the propagator with respect to the dimensionless component
$x_{\mathrm{in},\ell}/\Phi_0$ as
\begin{equation*}
V_\ell(t)\equiv
\left.\Phi_0\frac{\partial U(t;\bm{x}_{\mathrm{in}})}
{\partial x_{\mathrm{in},\ell}}\right|_{\bm{x}_{\mathrm{in}}=\bm 0}.
\end{equation*}
Let $H_0(t)=H_{\QCQ}[\bar\Phi(t)]$, and let $U_0(t)$ denote the corresponding
reference propagator. We also define
$H_\Phi(t)=\left.\partial_\Phi H_{\QCQ}\right|_{\bar\Phi(t)}$. Using the flux
sensitivity $h_\ell(t)=C\e^{At}\bm e_\ell$ from
Eq.~\eqref{eq:initial-state-flux-sensitivity}, differentiation of the
Schr\"odinger equation gives
\begin{equation}
\ii\dot V_\ell(t)
=H_0(t)V_\ell(t)+H_\Phi(t)h_\ell(t)U_0(t),
\qquad V_\ell(0)=0.
\label{eq:supp-propagator-tangent}
\end{equation}
Transforming Eq.~\eqref{eq:supp-propagator-tangent} to the interaction picture
and integrating from $0$ to $T$ gives
\begin{equation}
\begin{aligned}
U_0^\dagger(T)V_\ell(T)&=-\ii\mathcal G_\ell,\\
\mathcal G_\ell
&=\int_0^T h_\ell(t)
U_0^\dagger(t)H_\Phi(t)U_0(t)\,\dd t.
\end{aligned}
\label{eq:supp-interaction-frame-generator}
\end{equation}
Equation~\eqref{eq:supp-interaction-frame-generator} identifies
$\mathcal G_\ell$ as the first-order error generator. Its contribution to the
interaction picture propagator is $-\ii\mathcal G_\ell$. Expanding the propagator
in $x_{\mathrm{in},\ell}/\Phi_0$ therefore gives
\begin{equation}
U(T;x_{\mathrm{in},\ell})
=U_0(T)\left[\id-\ii\frac{x_{\mathrm{in},\ell}}{\Phi_0}
\mathcal G_\ell\right]
+\cO\!\left[\left(\frac{x_{\mathrm{in},\ell}}{\Phi_0}\right)^2\right],
\label{eq:supp-propagator-derivative}
\end{equation}
Therefore, first-order memory robustness requires
$\mathcal G_\ell^{\mathrm{log}}=0$ and
$\mathcal G_\ell^{\mathrm{leak}}=0$ for every retained component of the initial
actuator state, as stated in Eq.~\eqref{eq:memory-transparency-conditions}.

\section*{Data availability statement}

No experimental data were generated in this study.  The experimental step
response used to parameterize the three-pole flux line model was obtained from
Ref.~\cite{Li2025}.  The coefficients of the primary memory-robust pulse are
reported in Appendix~\ref{sec:appendix-optimization}.  Additional numerical data
supporting the findings of this study are available from the corresponding
author upon reasonable request.

\addcontentsline{toc}{chapter}{Acknowledgment}
\section*{Acknowledgment}

\begingroup
\raggedright
This work was supported by the Key-Area Research and Development Program of Guangdong Province (Grant No. 2018B030326001), the Science, Technology and Innovation Commission of Shenzhen Municipality (Grant Nos. JCYJ20170412152620376 and KYTDPT20181011104202253), the Innovation Program for Quantum Science and Technology (2021ZD0301703), Guangdong Major Project of Basic Research (2025B0303000007), the Shenzhen Science and Technology Program (Grant No. KQTD20200820113010023), and CCF-QuantumCtek Superconducting Quantum Computing Special Cooperation Program (No. CCF-QC2025002).
\par
\endgroup

\FloatBarrier
\addcontentsline{toc}{chapter}{References}
\begingroup
\footnotesize
\bibliographystyle{iopart-num}
\bibliography{references}
\endgroup

\end{document}